\documentclass[%
 reprint,
 amsmath,amssymb,
 aps,
 showkeys,
]{revtex4-2}

\usepackage{graphicx}
\usepackage{dcolumn}
\usepackage{bm}
\usepackage{xcolor}
\usepackage[T1]{fontenc}
\usepackage[normalem]{ulem}

\newcommand{\fp}{f^+(s)}
\newcommand{\fm}{f^-(s)}

\begin{document}

\preprint{APS/123-QED}

\title{Accessing Semi-Addressable Self Assembly with Efficient Structure Enumeration}

\author{Maximilian C. H\"ubl}
\email{maximilian.huebl@ist.ac.at}
\author{Carl P. Goodrich}%
\email{carl.goodrich@ist.ac.at}
\affiliation{%
 Institute of Science and Technology Austria (ISTA), Am Campus 1, 3400 Klosterneuburg, Austria
}%

\date{\today}

\begin{abstract}
Modern experimental methods enable the creation of self-assembly building blocks with tunable interactions, but optimally exploiting this tunability for the self-assembly of desired structures remains an important challenge.
Many studies of this inverse problem start with the so-called fully-addressable limit, where every particle in a target structure is different. This leads to clear design principles that often result in high assembly yield, but it is not a scaleable approach -- at some point, one must grapple with ``reusing'' building blocks, which lowers the degree of addressability and may cause a multitude of off-target structures to form, complicating the design process. 
Here, we solve a key obstacle preventing robust inverse design in the ``semi-addressable regime'' by developing a highly efficient algorithm that enumerates all structures that can be formed from a given set of building blocks. 
By combining this with established partition-function-based yield calculations, we show that it is almost always possible to find economical semi-addressable designs where the entropic gain from reusing building blocks outweighs the presence of off-target structures and even increases the yield of the target.
Thus, not only does our enumeration algorithm enable robust and scalable inverse design in the semi-addressable regime, our results demonstrate that it is possible to operate in this regime while maintaining the level of control often associated with full addressability. 
\end{abstract}


\maketitle
\begin{figure}
    \centering
    \includegraphics[width=\linewidth]{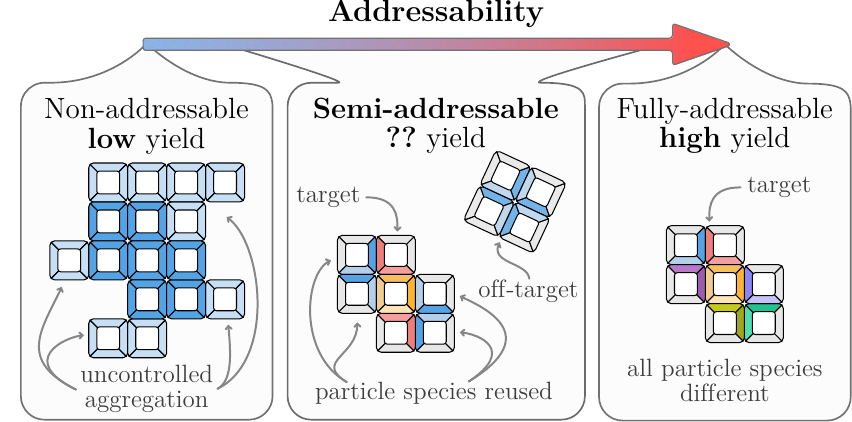}
    \caption{Different degrees of addressability. Non-addressable building blocks lead to uncontrolled aggregation, while a precise target shape can be encoded with fully-addressable building blocks. Depending on the degree of addressability, semi-addressable designs lead to varying numbers of competing structures, whose impact on yield is often hard to predict.}
    \label{fig:1}
\end{figure}

The so-called fully-addressable limit, where every particle in a target structure is different and individually tunable, is a central paradigm in inverse self assembly~\cite{hormoz2011, Zeravcic.20140h, Jacobs.2016, Jacobs.2015awj, Cademartiri.2015, Hedges.2014gmb}, applicable in a wide range of systems with programmable interactions~\cite{Wei.2012, Ke.2012, Park.2006, Sigl.2021, Hayakawa.2022pqs, Hayakawa.2024, Duque.2024, Evans.2024}. This limit leads to clear design principles for particle-particle interactions that prevent uncontrolled aggregation and allow the targeting of complex, precisely defined structures. In many systems, however, making every particle different is not scalable, as it couples (experimental) cost and complexity to the size and number of desired structures. To overcome this, one must venture away from the fully-addressable limit and into what we call the \emph{semi-addressable regime} (Fig.~\ref{fig:1}), where building blocks may fit together in more than one way.

In contrast to fully-addressable assembly, there exists an enormous number of semi-addressable designs for a given target, and most of these designs also lead to a large number of off-target structures that lower the chance of successful assembly.
On the other hand, reusing the same particle species multiple times in the same structure raises the structure's configurational entropy~\cite{Frenkel.2015, Whitelam.2015}, which positively affects its yield. 
Does this mean that it is possible to find semi-addressable designs with high target yields in spite of the off-target structures? Identifying such designs is a considerable challenge, as it is not enough to design \emph{for} a target structure, one must simultaneously design \emph{against} all the off-targets. 
Nevertheless, there is a path forward: if all competing structures are known, equilibrium yields can be efficiently predicted through an established partition-function-based approach~\cite{Holmes-Cerfon.2016, curatolo.2023, Klein.2018, Murugan.2015}. In many cases, this calculation is fast enough that one could iterate over the discrete binding rules (\textit{e.g.} using the approach of Ref.~\cite{Bohlin.2023}) and thus perform inverse design. 
The bottleneck with this approach is actually enumerating all the relevant off-target structures, the number of which can change by many orders of magnitude with very slight changes to the binding rules. 

In this paper, we overcome this major obstacle by introducing a general and highly efficient algorithm for enumerating self-assembled structures. This algorithm operates on minimal assumptions and can be used in many experimental or computational settings. 
We then show how this algorithm enables robust inverse design in the semi-addressable regime by combining it with partition-function-based yield calculations to identify nontrivial interaction patterns that maximize the yield of desired structures. 
Contrary to the natural expectation that off-target structures necessarily hurt the assembly of the target, we show that there almost always exists a semi-addressable design where this is overcome by the entropic benefits of fewer species, leading to better yields compared to the fully-addressable limit. This result means that yield does not have to be compromised to satisfy experimental constraints or to access other benefits of the semi-addressable regime (\emph{e.g.} designing reconfigurable structures, hierarchical structures~\cite{Whitelam.2015, Gruenwald.2014, Hayes.2021}, or multiple structures in parallel~\cite{Murugan.2015kjd, Sartori.2020, Osat.2023}).

\begin{figure}
    \centering
    \includegraphics[width=\linewidth]{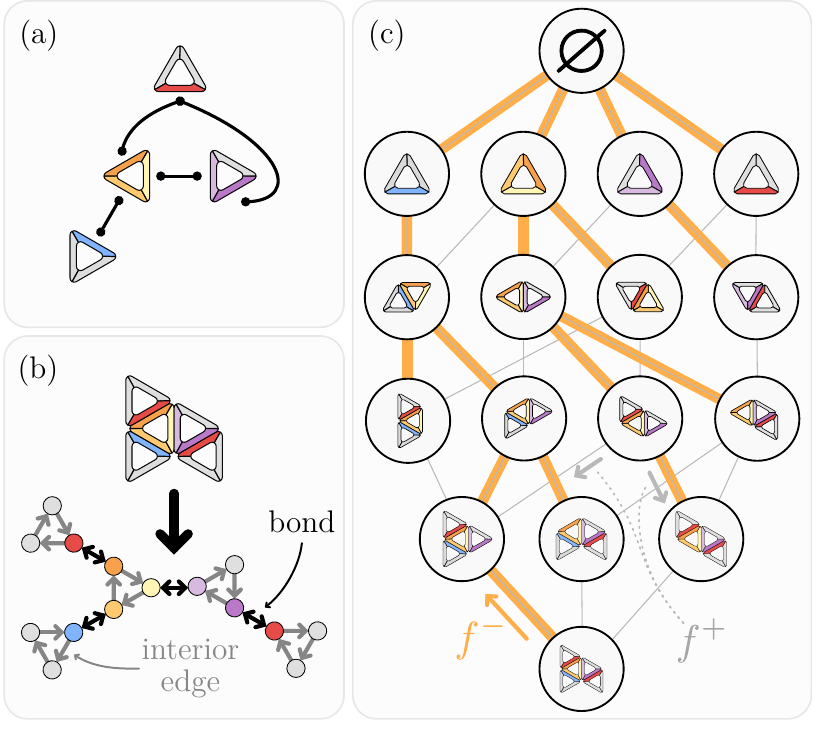}
    \caption{Reverse search enumeration. \textit{(a)} Diagrammatic representation of an assembly system consisting of 4 triangular building blocks with potential binding sites on the triangle edges. Two edges can bind only if they are connected in the diagram. 
    \textit{(b)} A structure and its graph encoding. \textit{(c)}
    Reverse search constructs a network of structures by adding particles with $\fp$ (grey lines) and removing particles with $\fm$ (orange lines).
    }
    \label{fig:2}
\end{figure}

We begin by defining an \emph{assembly system} as a set of building block species, each containing a number of programmable binding sites that follow specific binding rules, as sketched in Fig.~\ref{fig:2}a. Self assembly occurs when particles of each species type are mixed at finite concentrations, and particles are able to form discrete, orientationally-locking bonds \emph{only} if indicated by the binding rules (black lines in Fig.~\ref{fig:2}a).

To keep our algorithm broadly applicable, we make no further assumptions about the physical nature of the bonding interactions, but we note that programmable bonds of this type are common in DNA-based self-assembly, such as DNA-Origami particles~\cite{Sigl.2021, Hayakawa.2022pqs, Videbaek.2022, Duque.2024, Videbaek.2024,  Kahn.2022, Hayakawa.2024} or DNA-tiles~\cite{Park.2006, Wei.2012, Ke.2012, Evans.2024}, \emph{de-novo} proteins~\cite{Huang.2016, Boyken.2016}, and magnetic handshake materials~\cite{Niu.2019}.
While most real-world systems exhibit a finite amount of bond flexibility, which can lead to strained off-target assemblies~\cite{Hayakawa.2022pqs, Videbaek.2022, Videbaek.2024, Duque.2024}, we consider an idealized case of rigid assemblies where bonds are perfectly orientationally-locking, and we focus below only on off-target structures resulting from a lack of bond selectivity.
A possible generalization of our algorithm to include flexible bonds is discussed in the SI.

Given an assembly system and optional constraints (\emph{e.g.} maximum structure size), we use reverse search~\cite{Avis.1996, Avis.2021, Avis.2018} to efficiently enumerate the set of all structures $\mathcal{S}$ that can be formed by connecting building blocks site-to-site.
Reverse search is a general enumeration procedure that can be applied to a wide range of enumeration or search problems, and guarantees both a runtime linearly proportional to the number of structures $N_\mathrm{str} = |\mathcal{S}|$ and a memory requirement independent of $N_\mathrm{str}$~\cite{Avis.1996}.
In our context, reverse search works by generating new structures from pre-existing ones, through repeated addition and removal of particles.
More specifically, we employ a function $\fp$ that returns all structures that can be created by adding a particle to a pre-existing structure $s$, and a function $\fm$ that returns a single structure by removing one specific particle from $s$.

Importantly, reverse search requires that the order of the structures returned by $\fp$, while arbitrary, only depends on the topology of the structure, and not on the way the structure is represented numerically (\emph{e.g}. order or positions of the constituent particles). The same is true of the single (arbitrary) structure returned by $\fm$.
Making these functions well-defined is a key technical challenge, which we solve by encoding structures as vertex-colored directed graphs.
We represent binding sites as graph vertices, and indicate the structure topology with ``interior edges'' that connect binding sites belonging to the same particle, or ``bond edges'' that indicate a bond between particles (Fig.~\ref{fig:2}b).
By applying graph isomorphism tools~\cite{Babai.1983, McKay.2014} to this encoding, we can then consistently define the order in which particles are added or removed by $\fp$ and $\fm$.
See SI for implementation details and discussion.

With $\fp$ and $\fm$ in hand, reverse search works by generating structures through $\fp$ and then filtering them through $\fm$, as shown in Fig.~\ref{fig:2}c. Starting from the empty structure $s=\varnothing$, we use $\fp$ to add particles to $s$, generating tentative offspring structures $s'_i = \left[\fp\right]_i$.
However, we only accept those offspring that lead back to $s$ after removing a particle with $f^-(s'_i)$, \textit{i.e.} we accept only those $s'_i$ for which $f^-(s'_i) = s$.
This process is repeated for all accepted structures until no new offspring are found, at which point we have achieved a full enumeration. The properties of $\fp$ and $\fm$ discussed above guarantee that every structure is accepted exactly once.

\begin{figure*}
    \centering
    \includegraphics[width=\linewidth]{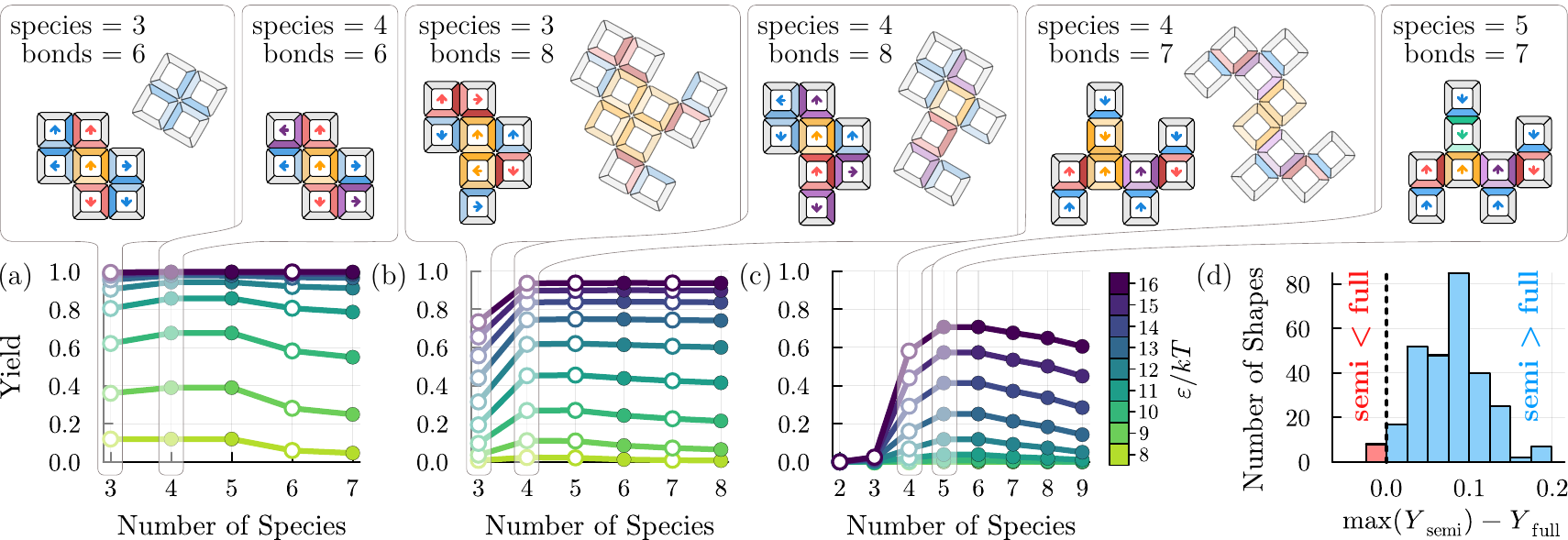}
    \caption{Inverse design in the semi-addressable regime. For a given target shape containing \textit{(a)} 7, \textit{(b)} 8, and \textit{(c)} 9 particles, we identify designs with a specified number of species that maximize equilibrium yield, which are shown at different binding energies $\varepsilon$ and volume fraction $\phi = 0.2$. 
    Only the highest-yield system for each species number are shown. Systems may be deterministic (full circles) or non-deterministic (empty circles) and the systems with maximal number of species are fully-addressable. Colored target shapes and some competing structures (greyed out) are shown for select systems, the arrows indicate the building block orientation. \textit{(d)} Histogram of the difference between the maximal semi-addressable yield $\max(Y_\mathrm{semi})$ and the fully-addressable yield $Y_\mathrm{full}$ for all square shapes of size $n\leq7$ at $\varepsilon = 14 \, kT$ and $\phi = 0.2$.}
    \label{fig:3}
\end{figure*}

A key strength of our approach is that it is easily adapted to any system with orientationally locking interactions, as our algorithm decouples the low-level enumeration procedure from the physics and geometry of the building blocks.
In contrast to other applications of reverse search to structure enumeration problems~\cite{Caporossi.1998, Liang.2017}, our graph-encoding technique allows our algorithm to work off-lattice, handle multiple species, and incorporate binding rules and constraints.
Building blocks that tile space or are convex can lead to performance gains, but are not required.
We find that the algorithm typically enumerates roughly 10,000 structures per second, and we easily perform enumerations of systems with tens of millions of structures.
Benchmark results and verification of the algorithm are shown in the SI.

We now return to the problem of semi-addressable inverse design.
Here, it is our goal to find economical assembly systems with as few building block species as possible that can still assemble a given target structure shape at high yield.
We start by considering three example target shapes: a 7-particle symmetric shape (Fig.~\ref{fig:3}a), an 8-particle asymmetric shape (Fig.~\ref{fig:3}b), and a 9-particle tree-like shape (Fig.~\ref{fig:3}c). For each shape, we perform a quasi-brute-force search~\footnote{For the structures considered here, this search took a few hours on a 2019 MacBook Pro.} over assembly systems to identify the design with the highest yield of the target shape at fixed $N_\mathrm{spc}$. 
To do this, we start from the fully-addressable assembly system corresponding to the target shape and then recursively merge building block species with each other, thereby generating increasingly economical designs with fewer and fewer species (see SI).
For convenience, we reject designs that allow more than $2000$ different structures to form.

For each design, we enumerate all structures that can possibly form, which allows us to compute equilibrium structure yields via the methods outlined in Refs.~\cite{Klein.2018, curatolo.2023, Murugan.2015}.
Briefly, the yield of a structure $s$ is $Y_s = \rho_s / \sum_{s^\prime \in \mathcal{S}} \rho_{s^\prime}$, where $\rho_s$ is the equilibrium number density
\begin{equation}\label{eq:rho}
    \rho_s = \frac{A_s}{\sigma_s} \, e^{\beta\left[\sum_\alpha n^\alpha_s \mu_\alpha + b_s \varepsilon \right]} \,,
\end{equation}
$\beta = 1/kT$ is the inverse temperature, $\mu_\alpha$ is the chemical potential of building block species $\alpha$, $\varepsilon$ is the binding  energy (which is assumed here to be the same for all bonds), $n^\alpha_s$ is the number of building blocks of species $\alpha$ in $s$, $b_s$ is the number of bonds in $s$ and $\sigma_s$ is the symmetry number of $s$.
$A_s$ is a prefactor that is related to the rotational and vibrational entropy of $s$, and depends on the details of the binding interactions. We ignore this prefactor in our initial search but include it when reporting final yields in Fig.~\ref{fig:3}.
Chemical potentials $\mu_\alpha$ are chosen to correspond to stoichiometric concentrations with total particle volume fraction $\phi = 0.2$. 
More details on the calculations can be found in the SI.

Once the yields are calculated, we find the design that maximizes the yield of the target shape for a given number of species $N_\mathrm{spc}$ -- Fig.~\ref{fig:3}a-c shows this maximum yield as a function of $N_\mathrm{spc}$ and binding energy $\varepsilon$. 
For all three target structure shapes, yield either increases or stays roughly constant as we decrease $N_\mathrm{spc}$ from the fully- into the semi-addressable regime.
Semi-addressable designs often exploit symmetries in the target structure (when applicable, \emph{e.g.} Fig.~\ref{fig:3}a), but optimal designs do much more than just exploit symmetries, as the shapes in Fig.~\ref{fig:3}b-c are not symmetric but still exhibit economical high-yield solutions. 
In these examples, optimal solutions sometimes exploit steric repulsion between the inert (grey) sides of the building blocks to `protect' open bonds, or even contain unprotected open bonds.
As highlighted by the open circles in Fig.~\ref{fig:3}, many of the more economical  systems are \emph{non-deterministic} in that they allow off-target structures that are not substructures of the target to form.

Figure~\ref{fig:3}a-c show optimal designs for three arbitrarily chosen target shapes, but these results are highly general. To demonstrate this, we performed the same calculation for all shapes composed of $n\leq7$ squares. As shown in Fig.~\ref{fig:3}d, at $\varepsilon = 14\,kT$ and $\phi = 0.2$ we find that for 276 out of 284 shapes (97\%), there exist semi-addressable designs that explicitly increase yield compared to the fully-addressable case. This demonstrates the generality of our results, implying that combining high yield with economy is a far-reaching and realistic design goal. 

\begin{figure}
    \centering
    \includegraphics[width=\linewidth]{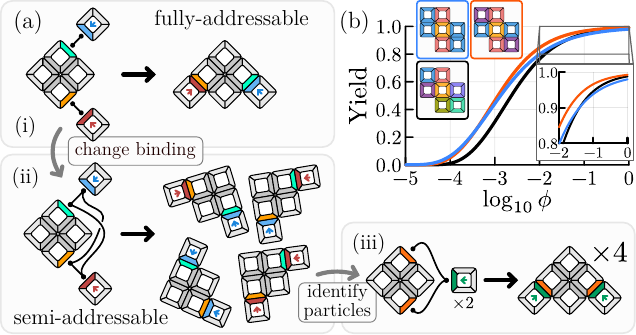}
    \caption{Configurational entropy gain. \textit{(a)} The fully-addressable design in (i) can be modified into a semi-addressable design that allows the formation of four version of the target structure (ii). Since the building blocks are now functionally identical, this system is completely equivalent to a simpler semi-addressable system, where the distinction between the two building blocks is dropped (iii).
    \textit{(b)} Yield of systems from Fig.~\ref{fig:3}a as a function of $\log_{10}\phi$, obtained at $\varepsilon=13\,kT$. Shown are the non-deterministic system with $N_\mathrm{spcs} = 3$ (blue), the deterministic system with $N_\mathrm{spcs} = 4$ (orange) and the fully-addressable system (black). Inset shows a zoomed in view of the top-right region of the plot. Note that the $\phi \to 1$ limit is generally inaccessible due to kinetic and/or steric constraints that are not considered here.}
    \label{fig:4}
\end{figure}

The observed yield increase in the semi-addressable regime comes from the configurational entropy gain associated with permutations of indistinguishable particles \cite{Whitelam.2015, Frenkel.2015}.
To understand this better, consider the fully-addressable design shown in Fig.~\ref{fig:4}a(i), where a red and a blue monomer act as ``caps'' and bind to a central region (four gray particles) as shown. 
The fully-addressable design can be modified so that the red and blue particles both bind to the teal and orange sites (Fig.~\ref{fig:4}a(ii)). 
This results in four possible structures, each with the same equilibrium number density as the original structure (assuming equal chemical potentials). 
Since the red and blue particles now have identical interactions, distinguishing between them has no physical meaning and they can be considered the same species (Fig.~\ref{fig:4}a(iii)).
In this case, the four structures in Fig.~\ref{fig:4}a(ii) become a single structure that appears at four times the number density, meaning we have constructed a semi-addressable design that exploits configurational entropy to enhance yield.

More generally, a fully-addressable design with $M$ such ``capping'' particles can be transformed into a semi-addressable design by combining the caps in this way. 
Every structure $s$ with $k_s$ caps has $M^{k_s}$ counterparts in the semi-addressable system, one for every permutation of cap species. When we stop distinguishing the caps, these counterparts become the \emph{same} structure, which now has extra configurational entropy, leading to a free energy reduction of
\begin{equation}
    \Delta F_s =  -\frac{k_s}{\beta} \log M \,.
\end{equation}

For equal chemical potentials in the fully- and semi-addressable systems, Eq.~\eqref{eq:rho} predicts that the number densities of these semi-addressable structures are $\rho^\mathrm{semi}_s = e^{-\beta\Delta F_s} \rho^\mathrm{full}_s$, where $\rho^\mathrm{full}_s$ are the number densities of the corresponding fully-addressable structures.
Since the target structure contains every cap, it has the largest configurational entropy gain and therefore its yield necessarily increases.

This illustrates the principle that reducing the number of species can increase configurational entropy and lead to larger yields. 
The same basic principle is at work in the systems shown in Fig.~\ref{fig:3}, except that many of those systems merge multiple groups of building blocks and are not restricted to caps (\emph{e.g.} the $N_\mathrm{spc}=3$ system in Fig.~\ref{fig:3}a merges a pair and a quartet of building blocks, leading to a free energy reduction of roughly $7\,kT$, or about 5 - 10\% of its total bond energy for the range of experimentally relevant binding energies considered in Fig.~\ref{fig:3}~\cite{Jacobs.2015awj}).
Note that the reduction in free energy can be viewed as decreasing the \emph{critical assembly concentration}~\cite{Hagan.2021}, and that the free energy change affects the volume fraction and stoichiometry of the semi-addressable systems.
Chemical potentials for the systems in Fig.~\ref{fig:3} have been adjusted to allow comparison at equal $\phi$, and a more general discussion of the free energy reduction and critical assembly concentration can be found in the SI.

The simple scenario in Fig.~\ref{fig:4}a is an example of a so-called deterministic assembly system, where the reduction of building block species does not introduce additional competing structures. However, especially at low $N_\mathrm{spc}$, deterministic designs may not always exist, meaning that semi-addressable designs lead to additional competing structures that negatively affect target yield. It is \emph{a priori} unclear whether the entropy gain can compensate for this, and in fact, for the vast majority of non-deterministic designs, the number of competing structures is enormous and target yield is significantly reduced.
Nevertheless, it is a key finding of this paper that, for every target shape we have investigated, there exist semi-addressable designs where either these two effects roughly balance, or where the entropy gain outright wins, resulting in increased target yield.

The balance between these two effects in general also depends on $\phi$ and $\varepsilon$.
Fig.~\ref{fig:4}b shows the yields of various systems from Fig.~\ref{fig:3}a as a function of $\phi$.
The deterministic system ($N_\mathrm{spc}=4$ system in Fig.~\ref{fig:3}a) results in higher yield compared to the fully-addressable one at all volume fractions, while the performance of non-deterministic systems ($N_\mathrm{spc}=3$ system in Fig.~\ref{fig:3}a) depends on $\phi$.
At low $\phi$, the additional off-target structures do not play a large role and since the non-deterministic system has the fewest species and largest entropy gain, it performs better than the other systems.
At high $\phi$ however, the additional competing structures start to outweigh the entropy gain, and the non-deterministic system performs slightly worse than the fully-addressable system.
Notice though that this yield decrease at high $\phi$ is very small, and for all target shapes we considered, we find economical designs with yield comparable to fully-addressable assembly at all volume fractions.

These results show that both assembly economy and quality can be improved by decreasing the number of species, thus moving from the fully-addressable into the semi-addressable assembly regime. 
While other authors have found semi-addressable designs that prevent off-target structures, for example by exploiting symmetry~\cite{Hayakawa.2024, Duque.2024, Kahn.2022} or through other computational approaches~\cite{Ahnert.2010, Bohlin.2023}, our results show that off-target structures do not have to be avoided to still achieve high equilibrium yield. This observation has profound implications for exploiting semi-addressability for multifarious self assembly, where a single set of building blocks assembles into multiple target structures simultaneously~\cite{Murugan.2015kjd, Sartori.2020, Bohlin.2023}, or for any use case where building blocks need to be reused.
Furthermore, we are able to achieve these high-yield designs while adhering to the experimentally relevant constraints of equal binding energies and stoichiometric concentrations. Additionally optimizing over binding energies and particle concentrations should lead to further increases in target yield~\cite{Murugan.2015}, but is not necessary for the structures we study. 

This result is enabled by our enumeration algorithm, which is so computationally efficient that we were able to perform inverse design through a quasi-brute-force approach (Fig.~\ref{fig:3}).
Combining our fast enumeration and yield calculations with more efficient methods for generating candidate assembly systems, such as the techniques of Bohlin et al.~\cite{Bohlin.2023}, should allow semi-addressable inverse design of larger and more complex structures. 
However, care must be taken when considering structures that self-close on length scales large compared to particle size, such as tubules~\cite{Hayakawa.2022pqs, Videbaek.2022, Videbaek.2024}, shells~\cite{Sigl.2021}, or other complex surfaces~\cite{Duque.2024}. 
In these cases, even small amounts of bond flexibility can lead to the formation of strained off-target structures that violate the assumption of perfectly rigid bonds, and suppressing such off-target structures generally requires sacrificing some assembly economy~\cite{Hayakawa.2022pqs, Videbaek.2022, Videbaek.2024, Duque.2024}.
Our algorithm could be modified in future work to include bond flexibility (see SI), which would extend our method to this more general class of systems.

Finally, we note that structure enumeration can be critical for addressing a wide range of topics in self assembly beyond the scope of this work.
For example, the structure network (Fig.~\ref{fig:2}c) could be exploited for kinetic calculations~\cite{Trubiano.2021, Perkett.2014, Gartner.2022, Gartner.2024}, or hierarchical assembly schemes~\cite{Whitelam.2015, Gruenwald.2014, Hayes.2021} could be implemented by reusing high-yield output structures as building blocks for a second round of self assembly.
We envision our enumeration algorithm, which is freely available online~\cite{Roly.jl}, to be a flexible tool for many such applications.

\begin{acknowledgements}
We thank Daichi Hayakawa, Thomas E. Videb{\ae}k, and W. Benjamin Rogers for important discussions, and J\'er\'emie Palacci, An{\dj}ela \v{S}ari\'c, and Scott Waitukaitis for helpful comments on the manuscript. The research was supported by the Gesellschaft f\"ur Forschungsf\"orderung Nieder\"osterreich under project FTI23-G-011. 
\end{acknowledgements}

\bibliography{refs.bib}
\end{document}